\documentclass[12pt]{article}
\usepackage{amsmath,cite}
\usepackage{amssymb}
\usepackage{color}
\usepackage{braket}
\usepackage[utf8]{inputenc}
\usepackage{hyperref} 
\usepackage{subcaption}
\usepackage{float}

\def \be {\begin{equation}}
\def \ee {\end{equation}}
\def \ba {\begin{aligned}}
\def \ea {\end{aligned}}
\def \bea {\begin{eqnarray}}
\def \eea {\end{eqnarray}}

\begin{document}

\begin{titlepage}
\begin{flushright}
May 2026
\end{flushright}
\vspace{0.5cm}
\begin{center}
{\large \bf TBA equations for  $SU(r+1)$ 
quantum Seiberg--Witten 
curve: \\ higher-order Mathieu equation 
}

\lineskip .75em
 \vskip 2.5cm
 {Feiyu Peng$^{a,}$\footnote{pengfy@gs.zzu.edu.cn} and Hongfei Shu$^{a,b,c,}$\footnote{shuphy124@gmail.com, shu@zzu.edu.cn}
 }
 \vskip 2.5em
  {\normalsize\it 
 $^{a}$Institute for Astrophysics, School of Physics,
 Zhengzhou University, Zhengzhou, Henan 450001, China\\
 $^{b}$Beijing Institute of Mathematical Sciences and Applications, Beijing, 101408, China\\
 $^{c}$Yau Mathematical Sciences Center, Tsinghua University, Beijing, 100084, China
 }
 \vskip 3.0em
\end{center}
\begin{abstract}
We develop the ODE/IM correspondence for the higher-order Mathieu equation arising from the quantum Seiberg--Witten curve of the pure $SU(r+1)$ ${\cal N}=2$ supersymmetric Yang-Mills theory. From the subdominant solutions, we construct the Q-/Y-systems and derive the corresponding TBA equations. The dependence of the moduli parameters is found to be encoded in the boundary conditions of the Y-functions at $\theta \to -\infty$. From these boundary data, we derive an analytic expression for the effective central charge, which also governs the subleading contribution in the large-$\theta$ expansion of the TBA equations. Finally, we compare the large-$\theta$ expansion of the Q-function derived from the TBA equations with that obtained from the WKB method, which yields analytic agreement at subleading order and precise numerical agreement at the higher-order corrections.

\end{abstract}
\end{titlepage}

\tableofcontents
\newpage

\section{Introduction}

 Integrable structures in Seiberg--Witten (SW) theory provide a precise framework for understanding the non-perturbative phenomena in supersymmetric gauge theories \cite{Seiberg:1994rs,Seiberg:1994aj}. Over the past decades, various correspondences between SW theory and the integrable model have been uncovered. The SW curve was found to be identified with the spectral curve of a classical integrable model, while the SW differential plays the role of the associated one-form differential  \cite{Gorsky:1995zq,Martinec:1995by,Donagi:1995cf}. 
 In particular, the pure $SU(r+1)$ SW theory is related to the $r+1$-particle periodic Toda chain. The corresponding integrable model will be quantized when we consider the Nekrasov--Shatashvili limit of the Omega deformation, where the non-zero Omega deformation parameter plays the role of Planck constant \cite{Nekrasov:2009rc,Nekrasov:2013xda}. In this limit, the SW curve is quantized to the Baxter TQ relation of the quantum integrable model. 
    
Another remarkable connection between the SW theory and integrable model arises from applying the ODE/IM correspondence \cite{Dorey:1998pt,Bazhanov:1998wj} to the quantum SW curves. In the case of the $A_r$ Argyres--Douglas (AD) theory \cite{Argyres:1995jj,Argyres:1995xn}, the quantum SW curve takes the form of a Schr\"odinger equation with degree $r+1$ polynomial potential \cite{Gaiotto:2014bza,Ito:2017ypt}.  For the monomial potential, the corresponding integrable model of $A_{2n}$ AD theory is found to be the minimal model $M_{2,2n+3}$ \cite{Ito:2017ypt}. Generalizing the ODE/IM correspondence to the Schr\"odinger equation with arbitrary polynomial potentials, one finds that the exact quantum periods of the corresponding AD theory satisfy the thermodynamic Bethe ansatz (TBA) equations of the quantum integrable model \cite{Ito:2018eon}, which provide a solution to the Riemann--Hilbert problem of the Borel resummed quantum period formulated in \cite{Voros83}. Remarkably, the same TBA equations also arise from the conformal limit of the non-linear integral equations describing the wall-crossing of the BPS particles in the Gaiotto--Moore--Neitzke (GMN) formalism \cite{Gaiotto:2010okc,Gaiotto:2009hg,Cecotti:2014zga}, which can be obtained in principle once the BPS spectrum in a given moduli chamber is known\footnote{Similar procedures have been applied to the Wilson loops and correlation functions in AdS/CFT correspondence \cite{Alday:2010vh,Hatsuda:2010cc,Caetano:2012ac,Ouyang:2022sje}.}. The TBA equations for the $D$-type, $(A_m, A_n)$-type AD theories and the supersymmetric quantum mechanics can be found in \cite{Ito:2019jio,Ito:2023cyz,Ito:2024wxw,Hollands:2019wbr,Ito:2021boh,Ito:2021sjo,Ito:2024nlt,Ito:2025yyx}. See \cite{Ito:2025pfo} for the review on these directions.

The quantum SW curve of pure $SU(2)$ SYM appears as the modified Mathieu equation \cite{Mironov:2009uv,Zenkevich:2011zx}. This equation is more complicated than the polynomial potential Schr\"odinger equation appearing in AD theory, because it has two irregular singular points. The TBA equations for the $SU(2)$ quantum SW curve have been studied in two approaches. From the BPS spectrum viewpoint, the TBA equations have been derived in the GMN approach \cite{Grassi:2019coc,Hao:2025azt}, which reproduce the resurgent properties of the quantum periods of the pure $SU(2)$ SYM in the NS limit. From the integrability viewpoint, the ODE/IM correspondence has been generalized to the Mathieu equation obtained by rescaling the moduli parameter in the quantum SW curve. This construction leads to the massless sinh-Gordon TBA equation \cite{unp-zamo,Fioravanti:2019vxi}, which will be referred to as the integrability TBA in this paper. The integrable structure is related to the Liouville theory at self-dual point. Restoring the moduli parameter, the integrability TBA decomposes into two TBA equations, which coincide with the GMN-type TBA in the strong coupling regime. Moreover, this structure has also been extended to the $SU(2)$ SYM with matter \cite{Grassi:2021wpw,Imaizumi:2021cxf,Fioravanti:2022bqf}. The Mathieu equation thus provides a unified framework in which quantum SW curves, exact WKB analysis and integrable models meet in a fully non-perturbative way.

  To further develop this unified framework, we study the quantum SW curve of pure $SU(r+1)$ SYM \cite{Mironov:2009dv,Popolitov:2010bz}\footnote{Related attempts in this direction can be found in \cite{Fioravanti:2019awr,Yan:2020kkb}.}, which can be rescaled to the higher-order Mathieu equation. The central objective of the present paper is to generalize the ODE/IM correspondence to the higher-order Mathieu equation and to derive the corresponding integrability TBA equations. Once the moduli parameters are restored, these TBA equations are expected to provide an exact non-perturbative representation of the quantum periods. 
  
  In this paper, we construct the Q-/Y-systems from the subdominant solutions of the higher-order Mathieu equation, and then convert them to the TBA equations. The dependence on the moduli parameters is encoded in the boundary conditions of the Y-functions as $\theta \to -\infty$. From these boundary data, we derive an analytic form for the effective central charge, which also governs the subleading large-$\theta$ behavior of the TBA equations. We further compare the large-$\theta$ expansion of the TBA equations with the WKB expansion of the quantum curve.  At the subleading order, the two approaches agree analytically through the effective central charge, while higher-order corrections are shown to match with high numerical precision.

This paper is organized as follows. In Section \ref{sec:higher-order Mathieu equaion}, 
we first rescale the $SU(r+1)$ quantum SW curve to the higher-order Mathieu equation.  We then construct the Q-/Y-system from the subdominant solutions, and convert them into the TBA equations. In Section \ref{sec:large theta from TBA}, we first compute the large-$\theta$ expansion of Q-/Y-functions by expanding the kernels. We then derive the boundary conditions at $\theta\to-\infty$ of the Y-functions and compute the effective central charge analytically. In Section \ref{sec:large theta from WKB}, we apply the WKB method to compute the large-$\theta$ expansion of the Q-function, which will be tested against the expansion obtained from the TBA side. 
In the last Section, we will conclude and discuss the future possible developments. 

\section{TBA from higher-order Mathieu equation}\label{sec:higher-order Mathieu equaion}

The pure $SU(r+1)$ SW curve of ${\cal N}=2$ SYM is given
by \cite{Klemm:1994qs,Argyres:1994xh}
\begin{equation}\label{eq:SW curve}
\frac{\bar{\Lambda}}{2}\left(e^{y}+e^{-y}\right)=p^{r+1}-\sum_{a=2}^{r+1}u_{a}p^{r+1-a},
\end{equation}
 whose SW differential leads to a symplectic form $d\lambda=idy\wedge dp$ on the $(y,p)$
phase space. 
The coefficients $u_a$ parametrize the Coulomb branch moduli, while $\bar{\Lambda}$ is the dynamical scale of the theory. The SW differential associated with this curve determines the low-energy effective dynamics through its period integrals. In the Nekrasov-Shatashvili (NS) limit of the Omega-background, the SW curve is quantized by $p\to -\epsilon\partial_{y}$:
\begin{align}
\label{eq:QSWSUr+1}
 & \left[(-1)^{r+2}\epsilon^{r+1}\partial_{y}^{r+1}-\sum_{a=2}^{r+1}(-1)^{r+2-a}u_{a}\epsilon^{r+1-a}\partial_{y}^{r+1-a}+\frac{\bar{\Lambda}}{2}(e^{y}+e^{-y})\right]\psi=0,
\end{align}
where $\epsilon$ is the nonzero $\Omega$-background parameter in NS limit. This equation is a higher-order generalization of the Schr\"{o}dinger equation, where $\epsilon$ plays the role of Planck constant. Reparameterizing the parameters by 
\begin{equation}
    \label{change variables}
(-1)^{r+2-a}u_{a}\epsilon^{-a}=p_{a},\quad u_{r+1}=P^{r+1}\epsilon^{r+1},\quad \bar{\Lambda}=e^{(r+1)\theta}\epsilon^{r+1},
\end{equation}
one obtains the higher-order Mathieu equation
\begin{equation}
    \label{ODE r+1 theta P}
\left[(-1)^{r+2}\partial_{y}^{r+1}-\sum_{a=2}^{r}p_a\partial_y^{r+1-a}+P^{r+1}+\frac{e^{(r+1)\theta}}{2}(e^{y}+e^{-y})\right]\psi=0.
\end{equation}
The central objective of the present paper is to develop the ODE/IM correspondence for this equation. In the following subsections, we first study the asymptotic behaviors of the subdominant solutions of the ODE \eqref{ODE r+1 theta P}, and introduce the Q-/Y-functions from them, which satisfy the Q-/Y-system. Furthermore, we will convert the Y-system to the TBA equations, whose effective central charge will also be introduced.

\subsection{Q-/Y-system}
Since the ODE \eqref{ODE r+1 theta P} has two irregular singularities, we evaluate the asymptotic subdominant solutions at ${\rm Re}(y)\to\infty$ and ${\rm Re}(y)\to-\infty$ respectively, which leads to
\begin{equation}\label{eq:psi-pos}
    \psi_{+}( y , \theta)\sim N_+ \frac{1}{\sqrt{ e^{r\theta+\frac{r}{r+1} y } }} \exp{\left( - (r+1)\left(\frac{1}{2}\right)^\frac{1}{r+1} e^{\theta+\frac{ y }{r+1}} \right)},\quad y\to \infty,
\end{equation}
and
\begin{equation}\label{eq:psi-neg}
\psi_{-}(y,\theta)\sim\begin{cases}
\frac{N_{-}}{\sqrt{e^{r\theta-\frac{r}{r+1}y}}}\exp\left(-(r+1)\left(\frac{1}{2}\right)^{\frac{1}{r+1}}e^{\theta-\frac{y}{r+1}}\right) & y\to-\infty,\qquad\ \ \text{odd }r\\
\frac{N_{-}}{\sqrt{e^{r\theta-\frac{r}{r+1}y}}}\exp\left(-(r+1)\left(\frac{1}{2}\right)^{\frac{1}{r+1}}e^{-\frac{i\pi}{r+1}}e^{\theta-\frac{y}{r+1}}\right) & y\to-\infty-i\pi,\ \text{even }r
\end{cases}.
\end{equation}
Using the following invariance of the ODE:
\begin{equation}
    \begin{aligned}
            \Omega_{+}: y\to y+i\pi,\quad \theta \to\theta +\frac{i\pi}{r+1},\quad
    \Omega_{-}:y\to y-i\pi,\quad \theta \to\theta +\frac{i\pi}{r+1},
    \end{aligned}
\end{equation}
we can generate the subdominant solutions in other strips of $y$:
\begin{equation}
    \begin{aligned}
        \psi_{+,a}:=&\Omega_{+}^{a}\psi_{+}=\psi_{+}\left(y+i\pi a,\theta+\frac{i\pi a}{r+1}\right),\\\psi_{-,a}:=&\Omega_{-}^{-a}\psi_{-}=\psi_{-}\left(y+i\pi a,\theta-\frac{i\pi a}{r+1}\right).
    \end{aligned}
\end{equation}
Note that $\Omega_-\psi_+=\psi_+,\  \Omega_+\psi_-=\psi_-$.

We fix the normalization by $W[\psi_{\pm,a},\psi_{\pm,a+1},\cdots,\psi_{\pm,a+r}]=1$, from which the normalization factors $N_+$, $N_-$ can be determined. 
$\psi_{-,a}$ can be expanded in terms of the basis $\{\psi_{+,a}\}$. For example, 
when $r$ is even, we expand $\psi_{-,a}$ in terms of $\psi_{+,-\frac{r}{2}+1}$, $\psi_{+,-\frac{r}{2}+2}$, $\cdots$, $\psi_{+,\frac{r}{2}+1}$,
\begin{equation}\label{eq:exp psink even r}
    \begin{aligned}
        \psi_{-,a}=W_{(a),-\frac{r}{2}+2,\cdots,\frac{r}{2}+1}\psi_{+,-\frac{r}{2}+1}+W_{-\frac{r}{2}+1,(a),\cdots,\frac{r}{2}+1}\psi_{+,-\frac{r}{2}+2}+\cdots+W_{-\frac{r}{2}+1,\cdots,\frac{r}{2},(a)}\psi_{+,\frac{r}{2}+1},
    \end{aligned}
\end{equation}
where we have used the following notation to represent the Wronskians, 
\begin{equation} 
    \begin{aligned}
        W_{-\frac{r}{2}+1,\cdots,b-1,(a),b+1,\cdots,\frac{r}{2}+1}:=W[\psi_{+,-\frac{r}{2}+1},\cdots,\psi_{+,b-1},\psi_{-,a},\psi_{+,b+1},\cdots,\psi_{+,\frac{r}{2}+1}].
    \end{aligned}
\end{equation}
Substituting the expansion of $\psi_{-,a}$ into the Wronskian of $\{\psi_{-,a}\}$, for example, $W[\psi_{-,0},\psi_{-,1}]$, one finds that this Wronskian can be expanded as 
\begin{equation}\label{eq:Q2 as coef}
    W[\psi_{-,0},\psi_{-,1}]=-\det\left(\begin{array}{cc}
W_{(0),-\frac{r}{2}+1,\cdots,\frac{r}{2}} & W_{(1),-\frac{r}{2}+1,\cdots,\frac{r}{2}}\\
W_{(0),-\frac{r}{2}+2,\cdots,\frac{r}{2}+1} & W_{(1),-\frac{r}{2}+2,\cdots,\frac{r}{2}+1}
\end{array}\right)W[\psi_{+,-\frac{r}{2}+1},\psi_{+,\frac{r}{2}+1}]+\cdots.
\end{equation}
There is a coefficient that can be written as a determinant whose adjacent columns can be related by a $\Omega_{-}^{-1}$, while its adjacent rows can be related by a $\Omega_{+}$. 
More generally, 
in the expansion of $W[\psi_{-,0},\cdots,\psi_{-,a}]$, 
there is always a coefficient which can be converted into the determinant of a $(a+1)\times(a+1)$ matrix follows the same pattern. The conversion procedure can refer to Appendix \ref{appendix normalization}. 
For odd $r$, one will find similar coefficients using $\psi_{+,-\frac{r-1}{2}},\psi_{+,-\frac{r-3}{2}},\cdots,\psi_{+,\frac{r+1}{2}}$ as the basis. 
We then introduce the Q-function to describe the connection between the solutions $\psi_{\pm,a}$. From the structure of \eqref{eq:Q2 as coef}, we define the Q-function by the determinant 
\begin{equation}
    \label{Q_k+1 def}
\begin{aligned}
Q_{a+1}:=\pm\det\begin{pmatrix}-Q_{1} & -Q_{1}^{[-1]} & \cdots & -Q_{1}^{[-a]}\\
-Q_{1}^{[+1]} & -Q_{1} & \cdots & -Q_{1}^{[-a+1]}\\
\vdots & \vdots & \ddots & \vdots\\
-Q_{1}^{[+a]} & -Q_{1}^{[+a-1]} & \cdots & -Q_{1}
\end{pmatrix}.
\end{aligned}
\end{equation}
Here the sign depends on the value of $(a+1)$, which is negative when $a+1=4n+1$ or $a+1=4n+2$, and positive when $a+1=4n+3$ or $a+1=4n+4$. The shifted Q-functions are defined by $Q_{a}^{[j]}:=Q_a\left(\theta+\frac{i\pi j}{r+1}\right)$, and $Q_1$ is defined by the Wronskian 
\begin{equation}
    Q_{1}:=\begin{cases}
-W[\psi_{-,0},\psi_{+,-\frac{r-1}{2}},\cdots,\psi_{+,\frac{r-1}{2}}] & \text{odd }r\\
-W[\psi_{-,0},\psi_{+,-\frac{r-2}{2}},\cdots,\psi_{+,\frac{r}{2}}] & \text{even }r
\end{cases}.
\end{equation}
Using the Jacobi identity\footnote{
Here we use the Jacobi identity, 
\begin{equation}
    \Delta\,\Delta[1,a+1|1,a+1]=\Delta[a+1|a+1]\Delta[1|1]-\Delta[1|a+1]\Delta[a+1|1],
\end{equation}
where $\Delta$ represents the  determinant of an $(a+1)\times(a+1)$ matrix and $\Delta[a_1,a_2\dots|b_1,b_2\dots]$ represents the minor obtained from $\Delta$ by deleting $a_1,a_2\dots$-th rows and $b_1,b_2\dots$-th columns.
}, we find the Q-functions 
satisfy the Q-system
\begin{equation}
    \label{Q sys}
    Q_{a}^{[+1]}Q_{a}^{[-1]}=Q_{a}^{2}+Q_{a+1}Q_{a-1},\quad a=1,2\cdots r,
\end{equation}
where $Q_0=1$, and when the normalization is fixed by $W[\psi_{\pm,k},\psi_{\pm,k+1},\cdots,\psi_{\pm,k+r}]=1$, we find $Q_{r+1}=\pm1$, where the sign depends on the choice in \eqref{Q_k+1 def}. See Appendix \ref{appendix normalization} for more details. Also by defining $Y_{a}=Q_{a}^{2}/(Q_{a+1}Q_{a-1})$, the Q-system leads to the Y-system, 
\begin{equation}
    \label{Y sys general k}
    Y_{a}^{[+1]}Y_{a}^{[-1]}=\frac{(Y_{a}+1)^{2}}{(Y_{a+1}^{-1}+1)(Y_{a-1}^{-1}+1)}.
\end{equation}
The boundary conditions for the Y-system are $Y_{0}^{-1}=Y_{r+1}^{-1}=0$, which are equivalent to 
$Q_{-1}=Q_{r+2}=0$. When $r=1$, \eqref{Y sys general k} reproduces the previous results of \cite{unp-zamo, Fioravanti:2019vxi}.

\subsection{TBA equations}
We now convert the Y-system into the TBA equations. 
According to the Y-system, the pseudoenergy $\varepsilon_{a}(\theta):=-\log Y_{a}(\theta)$ satisfies
\begin{equation}\label{eq:eps-rel}
\varepsilon_{a}\left(\theta+\frac{i\pi}{r+1}\right)+\varepsilon_{a}\left(\theta-\frac{i\pi}{r+1}\right)=\varepsilon_{a+1}(\theta)+\varepsilon_{a-1}(\theta)-2L_{a}(\theta)+L_{a+1}(\theta)+L_{a-1}(\theta),
\end{equation}
where $L_{a}(\theta):=\log (e^{-\varepsilon_{a}(\theta)}+1)$. It will be explained later that the pseudoenergy behaves as $\varepsilon_{a}(\theta)\sim m_{a}e^{\theta}$ with positive $m_a$ at large $\theta $, which can be obtained by using the WKB method. 
Substituting this into \eqref{eq:eps-rel}, we find the constraints on $m_k$, 
\begin{equation}\label{eq:mk constraints}
m_{a}\left(e^{\frac{i\pi}{r+1}}+e^{-\frac{i\pi}{r+1}}\right)=m_{a+1}+m_{a-1}
\end{equation}
with $m_0=m_{r+1}=0$. Here we have used the fact that $L_k(\theta)$ approaches $0$ at large $\theta$. 
Defining $f_a(\theta):= \varepsilon_a(\theta)-m_a e^{\theta}$, which allows us to perform the Fourier transform and inverse Fourier transform\footnote{See \cite{Dorey:2007zx} for more details, where the Fourier transform is 
defined by $\tilde{f}(\omega):=\int_{-\infty}^{\infty}f(\theta)e^{-i\omega\theta}d\theta$. 
}, 
the equation \eqref{eq:eps-rel} leads to the TBA equations, 
\begin{equation}\label{eq:TBA-P}
    \varepsilon_{a}(\theta)=m_{a} e^{\theta}-\sum_{b=1}^{r}\int_{-\infty}^{\infty}\phi_{ab}(\theta-\theta^\prime)L_b(\theta^\prime)d\theta^\prime,
\end{equation}
where the kernel matrix $\phi$ is obtained from the inverse Fourier transform of $\tilde{\phi}(\omega )=-A^{-1}(\omega )B$. Here the matrices 
$A(\omega)$ and $B$ are defined by $A(\omega)=2\mathbb{I}_r\cosh\left(\frac{\omega\pi}{r+1}\right)-G_r$, $B=-C_r$, where $\mathbb{I}_r$ is the identity matrix, $G_r$ and $C_r$ are the incidence matrix\footnote{
Here the incidence matrix is defined by $(G_{r})_{ab}=\delta_{a,b+1}+\delta_{a,b-1}$.
} and the Cartan matrix respectively of the $A_r$ Dynkin diagram.
The kernel $\phi_{ab}$ for general $r$ reads 
\begin{equation}\label{eq: kernel phi}
    \phi_{ab}(\theta)=\frac{2}{\pi}\sum_{n=1}^{r}\sin\frac{n\pi a}{r+1}\sin\frac{n\pi b}{r+1}\tan\frac{n\pi}{2(r+1)}\frac{\sinh\left((r+1-n)\theta\right)}{\sinh\left((r+1)\theta\right)},
\end{equation}
and it satisfies 
\begin{equation}\label{eq:phitilde}
    \int d\theta\phi_{ab}(\theta)=\tilde{\phi}_{ab}(\omega=0)=\delta_{ab},\quad \int d\theta\theta\phi_{ab}(\theta)=i\frac{d}{d\omega}\tilde{\phi}(\omega)\Big|_{\omega=0}=0.
\end{equation}

The nonlinear integral equations for Q-functions can be obtained by rewriting the Q-system as
\begin{equation}\label{eq:Q-sys trans}
    Q_{a}^{[+1]}Q_{a}^{[-1]}=Q_{a+1}Q_{a-1}(1+Y_{a}).
\end{equation}
In analogy with the pseudoenergy, the logarithm of the Q-functions behaves as $\log Q_{a}\sim -q_{a}e^{\theta}$ at large $\theta $. According to the Q-system \eqref{eq:Q-sys trans}, we find the leading order contribution of $\log Q_a$ satisfies 
\begin{equation}\label{eq:qk constarins}
    q_{a}\left(e^{\frac{i\pi}{r+1}}+e^{-\frac{i\pi}{r+1}}\right)=q_{a+1}+q_{a-1}, 
\end{equation}
with $q_{0}=q_{r+1}=0$. 
Meanwhile, according to the definition of the Y-function, the constraints between $q_a$ and $m_a$ are given by
\begin{equation}\label{eq:mk qk constraints}
    m_{a}=2q_{a}-q_{a+1}-q_{a-1}. 
\end{equation}
Following a similar procedure used to derive the TBA equations, the Q-system \eqref{eq:Q-sys trans} can be converted to the nonlinear integral equations 
\begin{equation}\label{eq:TBA Q}
    -\log Q_{a}=q_{a}e^{\theta}-\sum_{b=1}^{r}\int d\theta^{\prime}\varphi_{ab}(\theta-\theta^{\prime})L_{b}(\theta^{\prime}),
\end{equation}
where the kernel matrix $\varphi$ is the inverse Fourier transform of $\tilde{\varphi}=A^{-1}(\omega)$, and the elements of the $\varphi$ matrix can be expressed as 
\begin{equation}\label{eq: kernel varphi}
    \varphi_{ab}(\theta)=\frac{1}{\pi}\sum_{k=1}^{r}\frac{\sin\left(\frac{ak\pi}{r+1}\right)\sin\left(\frac{bk\pi}{r+1}\right)}{\sin\left(\frac{k\pi}{r+1}\right)}\frac{\sinh((r+1-k)\theta)}{\sinh((r+1)\theta)}.
\end{equation}

\subsection{Boundary conditions and effective central charge} 
Note that the TBA equations are derived originally from the solutions of the higher-order Mathieu equation. However, the moduli parameters ($P$, $p_a$) do not appear explicitly in the TBA equations. 
 Their dependence is instead encoded in the boundary condition of $\varepsilon_a(\theta)$. At $\theta\to -\infty$, we impose the boundary condition
\begin{equation}\label{eq:hk-eps-bdy}
    \varepsilon_{a}(\theta)=h_{a}\theta+2C_{a}+\cdots,\quad\theta\to-\infty, 
\end{equation}
where $h_a$ are positive constants depending on $P$, $p_a$. This boundary condition is consistent with the TBA equations \eqref{eq:TBA-P} by noting \eqref{eq:phitilde}. As we will explain in Section \ref{chap:boundary condition}, the boundary condition can be obtained from the ODE \eqref{ODE r+1 theta P}. 
To impose the boundary condition, we use Zamolodchikov's trick to modify the TBA equations by 
\begin{equation}
    \varepsilon_{a}(\theta)=m_{a}e^{\theta}-f_{0,a}-\sum_{b=1}^{r}\int_{-\infty}^{\infty}\phi_{ab}(\theta-\theta^{\prime})\Big(L_{b}(\theta^{\prime})-L_{0,b}(\theta^{\prime})\Big)d\theta^{\prime}
\end{equation}
where $L_{0,b}:=\frac{h_{b}}{2}\log\big(1+e^{-2\theta}\big)$ and $f_{0,a}$ is given by 
\begin{equation}
    f_{0,a}=\sum_{b=1}^{r}\int_{-\infty}^{\infty}\phi_{ab}(\theta-\theta^{\prime})L_{0,b}(\theta^{\prime})d\theta^{\prime}.
\end{equation}
 Note that $f_{0,a}$, which depends on $P$ and $p_a$, approaches $-h_a \theta$ at $\theta\to-\infty$.

In the context of integrable model, the effective central charge for TBA equations \eqref{eq:TBA-P} is given by
\begin{equation}\label{eq:ceff}
    c_{{\rm eff}}=\frac{6}{\pi^{2}}\sum_{a=1}^{r}\int_{-\infty}^{\infty}d\theta m_{a}e^{\theta}L_{a}(\theta).
\end{equation}
Usually, the effective central charge can be computed using the standard dilogarithm method. 
However, due to the nontrivial boundary condition of $\varepsilon_a(\theta)$ at $\theta\to-\infty$, the method has to be modified, which leads to, 
\begin{equation}
    \begin{aligned}
        c_{{\rm eff}}=&\frac{6}{\pi^{2}}\sum_{a}{\cal L}(\frac{1}{1+e^{\varepsilon_{a}(-\infty)}})+\frac{3}{2\pi^{2}}\sum_{a,b}h_{a}h_{b}\int_{-\infty}^{\infty}\phi_{ab}(\theta)\theta^{2}d\theta\\
        =&r
        -\frac{3}{2\pi^{2}}\sum_{a,b}h_{a}h_{b}\frac{d^{2}}{d\omega^{2}}\tilde{\phi}_{ab}(\omega)\Big|_{\omega=0},
    \end{aligned}
\end{equation}
where ${\cal L}$ is the dilogarithm function defined by
\begin{equation}
    {\cal L}(x)=-\frac{1}{2}\int_{0}^{x}dt\Big(\frac{\log t}{1-t}+\frac{\log(1-t)}{t}\Big).
\end{equation}
The effective central charges for $r=1,2,3$ are 
\begin{equation}\label{eq:ceff-with-bdy}
    c_{{\rm eff}}=                                                                                                                                  \begin{cases}
1+\frac{3}{8}h_{1}^{2} & r=1\\
2+\frac{2}{9}(h_{1}^{2}+h_{1}h_{2}+h_{2}^{2}) & r=2\\

3+\frac{3}{64}\left(3h_{1}^{2}+4h_{2}h_{1}+2h_{3}h_{1}+4h_{2}^{2}+3h_{3}^{2}+4h_{2}h_{3}\right) 
& r=3
\end{cases}.
\end{equation}

The large-$\theta$ asymptotic expansion of the $\varepsilon_a$ and $\log Q_a$ can be obtained by expanding the kernels $\phi$ and $\varphi$. On the other hand, the asymptotic behaviors of the Y-functions and the Q-functions can also be computed by the WKB method, where the $e^{-\theta}$ plays the role of the Plank constant. 
In the following sections, we will illustrate these methods and compare them. 

\section{Large-$\theta$ expansion from TBA side}\label{sec:large theta from TBA}
In this section, we perform the large-$\theta$ expansion of $\varepsilon_a(\theta)$ and $\log Q_a$ by expanding the kernels in the corresponding equations. The cases of $r=2,3$ will be illustrated in detail.  We then derive the boundary condition of $\varepsilon_a (\theta)$ at $\theta\to-\infty$, and compute the effective central charge analytically.

\subsection{Large-$\theta $ expansion of $\varepsilon_a(\theta)$}
At $\theta\to \infty$, we can expand the kernel of the TBA equation \eqref{eq:TBA-P}, which leads to
\begin{equation}\label{eq:eps-exp}
    \begin{aligned}
        \varepsilon_{a}(\theta)=m_{a}e^{\theta}+\sum_{n=1}^{\infty}m_{a}^{(n)}e^{-n\theta}.
    \end{aligned}
\end{equation}
Since the expansion of the kernel differs for different $r$, we will illustrate it for the cases of $r=2,3$. 

\paragraph{$r=2$} 
The matrix of kernel $\phi$ is given by 
\begin{equation}
    \phi(\theta)=\begin{pmatrix}\frac{\sqrt{3}}{2\pi}\frac{2\cosh(\theta)+3}{2\cosh(2\theta)+1} & \frac{\sqrt{3}}{2\pi}\frac{2\cosh(\theta)-3}{2\cosh(2\theta)+1}\\
\frac{\sqrt{3}}{2\pi}\frac{2\cosh(\theta)-3}{2\cosh(2\theta)+1} & \frac{\sqrt{3}}{2\pi}\frac{2\cosh(\theta)+3}{2\cosh(2\theta)+1}
\end{pmatrix},
\end{equation}
where $\phi_{11}=\phi_{22}$, $\phi_{12}=\phi_{21}$. Expanding them at $\theta\to\infty$, we find 
\begin{equation}
    \begin{aligned}
        m_{1}^{(1)}=&m_{2}^{(1)}=-\frac{\text{\ensuremath{\sqrt{3}}}}{2\pi}\int_{-\infty}^{\infty}e^{\theta}\Big(L_{1}(\theta)+L_{2}(\theta)\Big)d\theta,\\
        m_{1}^{(2)}=&-m_{2}^{(2)}=-\frac{3\sqrt{3}}{2\pi}\int_{-\infty}^{\infty}e^{2\theta}\Big(L_{1}(\theta)-L_{2}(\theta)\Big)d\theta,\\
        \cdots&. 
    \end{aligned}
\end{equation}
It is interesting to note that the subleading contribution $m_{1}^{(1)}$ is related to the effective central charge \eqref{eq:ceff} by
\begin{equation}\label{eq:m11 ceff r=2}
    m_{1}^{(1)}=-\frac{\pi\sqrt{3}}{12m_{1}}c_{{\rm eff}},
\end{equation}
where we have used the constraints \eqref{eq:mk constraints} on the masses, i.e. $m_1=m_2$ for $r=2$.

\paragraph{$r=3$}

The matrix of kernel is given by
\begin{equation}
    \phi(\theta)=\begin{pmatrix}\frac{2\cosh(\theta)+\left(\sqrt{2}-1\right)\cosh(2\theta)+\sqrt{2}}{\pi(\cosh(\theta)+\cosh(3\theta))} & -\frac{\left(\sqrt{2}-2\right)\cosh(2\theta)+\sqrt{2}}{\pi(\cosh(\theta)+\cosh(3\theta))} & \frac{-2\cosh(\theta)+\left(\sqrt{2}-1\right)\cosh(2\theta)+\sqrt{2}}{\pi(\cosh(\theta)+\cosh(3\theta))}\\
-\frac{\left(\sqrt{2}-2\right)\cosh(2\theta)+\sqrt{2}}{\pi(\cosh(\theta)+\cosh(3\theta))} & \frac{2\left(\left(\sqrt{2}-1\right)\cosh(2\theta)+\sqrt{2}\right)}{\pi(\cosh(\theta)+\cosh(3\theta))} & -\frac{\left(\sqrt{2}-2\right)\cosh(2\theta)+\sqrt{2}}{\pi(\cosh(\theta)+\cosh(3\theta))}\\
\frac{-2\cosh(\theta)+\left(\sqrt{2}-1\right)\cosh(2\theta)+\sqrt{2}}{\pi(\cosh(\theta)+\cosh(3\theta))} & -\frac{\left(\sqrt{2}-2\right)\cosh(2\theta)+\sqrt{2}}{\pi(\cosh(\theta)+\cosh(3\theta))} & \frac{2\cosh(\theta)+\left(\sqrt{2}-1\right)\cosh(2\theta)+\sqrt{2}}{\pi(\cosh(\theta)+\cosh(3\theta))}
\end{pmatrix},
\end{equation}
where the independent elements are $\phi_{11}$, $\phi_{12}$, $\phi_{13}$, $\phi_{21}$, $\phi_{22}$. 
Expanding the kernel at $\theta\to\infty$, we find at the order $e^{-\theta}$, 
\begin{equation}
    m_{1}^{(1)}=m_{3}^{(1)}=\frac{1}{\sqrt{2}}m_{2}^{(1)}=-\frac{\left(\sqrt{2}-1\right)}{\pi}\int_{-\infty}^{\infty}e^{\theta}\left(L_{1}(\theta)+\sqrt{2}L_{2}(\theta)+L_{3}(\theta)\right)d\theta. 
\end{equation}
And at the order $e^{-2\theta}$, 
\begin{equation}
    m_{1}^{(2)}=-m_{3}^{(2)}=-\frac{2}{\pi}\int_{-\infty}^{\infty}e^{2\theta}\left(L_{1}(\theta)-L_{3}(\theta)\right) d\theta,\quad m_{2}^{(2)}=0.
\end{equation}
$m_{1}^{(1)}$ is related to the effective central charge \eqref{eq:ceff} by 
\begin{equation}\label{eq:m11 ceff r=3}
    m_{1}^{(1)}=-\frac{\left(\sqrt{2}-1\right)\pi}{6m_{1}}c_{{\rm eff}},
\end{equation}
where we have also used \eqref{eq:mk constraints}, i.e. $\sqrt{2}m_{1}=\sqrt{2}m_{3}=m_{2}$ for $r=3$.

\subsection{
Large-$\theta$ expansion of the Q-function}

Expanding the kernel $\varphi$ in the integral equations for Q-functions \eqref{eq:TBA Q}, we obtain the large-$\theta$ expansion of $\log Q_a$, 
\begin{equation}\label{eq:Q-exp-r2-TBA}
    -\log Q_{a}=q_{a}e^{\theta}+\sum_{n=1}^{\infty}q_{a}^{(n)}e^{-n\theta}.
\end{equation}
Similarly, we can find how $q_1^{(1)}$ is related to the effective central charge. 
We will illustrate the expansions for the cases of $r=2$ and $r=3$.

\paragraph{$r=2$} 
The matrix $\varphi$ reads 
\begin{equation}
    \varphi=\left(\begin{array}{cc}
\frac{\sqrt{3}}{2\pi}\frac{1}{2\cosh\theta-1} & \frac{\sqrt{3}}{2\pi}\frac{1}{2\cosh\theta+1}\\
\frac{\sqrt{3}}{2\pi}\frac{1}{2\cosh\theta+1} & \frac{\sqrt{3}}{2\pi}\frac{1}{2\cosh\theta-1}
\end{array}\right).
\end{equation}
We expand the kernel at large $\theta$. At the order $e^{-\theta}$, $q_1^{(1)}$ and $q_2^{(1)}$ are related to the effective central charge \eqref{eq:ceff} as
\begin{equation}\label{eq:q1-ceff}
    q_{1}^{(1)}=q_{2}^{(1)}=-\frac{\sqrt{3}}{2\pi}\int_{-\infty}^{\infty}e^{\theta}\left(L_{1}(\theta)+L_{2}(\theta)\right)d\theta=-\frac{\pi\sqrt{3}}{12m_{1}}c_{{\rm eff}},
\end{equation}
where $m_1=m_2$ is used. 
And at the order $e^{-2\theta}$, the coefficients read 
\begin{equation}\label{eq:q12-r2-TBA}
    q_{1}^{(2)}=-q_{2}^{(2)}=-\frac{\sqrt{3}}{2\pi}\int_{-\infty}^{\infty}e^{2\theta}\left(L_{1}(\theta)-L_{2}(\theta)\right)d\theta.
\end{equation}

\paragraph{$r=3$}
The matrix $\varphi$ is given by
\begin{equation}
    \varphi=\left(\begin{array}{ccc}
\frac{\left(\sqrt{2}\cosh(\theta)+1\right)}{2\pi\cosh(2\theta)} & \frac{1}{2\pi\cosh(\theta)} & \frac{\left(\sqrt{2}\cosh(\theta)-1\right)}{2\pi\cosh(2\theta)}\\
\frac{1}{2\pi\cosh(\theta)} & \frac{\sqrt{2}\cosh(\theta)}{\pi\cosh(2\theta)} & \frac{1}{2\pi\cosh(\theta)}\\
\frac{\left(\sqrt{2}\cosh(\theta)-1\right)}{2\pi\cosh(2\theta)} & \frac{1}{2\pi\cosh(\theta)} & \frac{\left(\sqrt{2}\cosh(\theta)+1\right)}{2\pi\cosh(2\theta)}
\end{array}\right).
\end{equation}
Expanding $\varphi$ at large $\theta$, the first several coefficients read 
\begin{equation}\label{eq:q12-r3-TBA}
    \begin{aligned}
        q_{1}^{(1)}=&q_{3}^{(1)}=\frac{1}{\sqrt{2}}q_{2}^{(1)}=-\frac{1}{\sqrt{2}\pi}\int_{-\infty}^{\infty}e^{\theta}\left(L_{1}(\theta)+\sqrt{2}L_{2}(\theta)+L_{3}(\theta)\right)d\theta,\\q_{1}^{(2)}=&-q_{3}^{(2)}=-\frac{1}{\pi}\int_{-\infty}^{\infty}e^{2\theta}\left(L_{1}(\theta)-L_{3}(\theta)\right)d\theta,\qquad q_{2}^{(2)}=0.
    \end{aligned}
\end{equation}
The subleading contribution $q_1^{(1)}$ is related to the effective central charge \eqref{eq:ceff} as
\begin{equation}\label{eq:q1-ceff r3}
    \begin{aligned}
q_{1}^{(1)}=-\frac{\pi}{6\sqrt{2}m_{1}}c_{{\rm eff}},
    \end{aligned}
\end{equation}
where we have used $\sqrt{2}m_{1}=\sqrt{2}m_{3}=m_{2}$.

\subsection{Boundary condition at $\theta\to -\infty$}
\label{chap:boundary condition}
We then study the boundary condition of Q-/Y-function at $\theta \to -\infty$, where the solution of the ODE \eqref{ODE r+1 theta P} can be expressed as
\begin{equation}
    \psi_{\lambda_i}\sim e^{\lambda_{i}y},\ \theta\to-\infty.
\end{equation}
Here $\lambda_i$ are the solutions of  $(-1)^{r+2}\lambda_{i}^{r+1}-\sum_{a=2}^{r}p_{a}\lambda_{i}^{r+1-a}+P^{r+1}=0$ ordered by $\mathrm{Re}(\lambda_{1})>\mathrm{Re}(\lambda_{2})>\cdots>\mathrm{Re}(\lambda_{r+1})$, which satisfy $\sum_{i=1}^{r+1}\lambda_i=0$.  
Using the basis $\{\psi_{\lambda_i}\}$, $\psi_{-,0}$ and $\psi_{+,a}$ can be expanded by 
\begin{equation}
    \begin{aligned}
       \psi_{-,0}=&\sum_{m=1}^{r+1}W[\psi_{\lambda_{1}},\cdots,\psi_{\lambda_{m-1}},\psi_{-,0},\psi_{\lambda_{m+1}},\cdots,\psi_{\lambda_{r+1}}]\psi_{\lambda_{m}},\\\psi_{+,a}=&\sum_{m=1}^{r+1}W[\psi_{\lambda_{1}},\cdots,\psi_{\lambda_{m-1}},\psi_{+,a},\psi_{\lambda_{m+1}},\cdots,\psi_{\lambda_{r+1}}]\psi_{\lambda_{m}}. 
    \end{aligned}
\end{equation}
According to the properties of the coefficients under $\Omega_\pm$, the coefficients can be expressed as
\begin{equation}
    \begin{aligned}
        W[\psi_{\lambda_{1}},\cdots,\psi_{\lambda_{m-1}},\psi_{-,0},\psi_{\lambda_{m+1}},\cdots,\psi_{\lambda_{r+1}}]=&e^{(r+1)(\sum_{i}\lambda_{i}-\lambda_{m})\theta}A_{-,i}(e^{2(r+1)\theta}),\\W[\psi_{\lambda_{1}},\cdots,\psi_{\lambda_{m-1}},\psi_{+,a},\psi_{\lambda_{m+1}},\cdots,\psi_{\lambda_{r+1}}]=&e^{-(r+1)(\sum_{i}\lambda_{i}-\lambda_{m})\left(\theta+\frac{2a\pi i}{r+1}\right)}A_{+,i}(e^{2(r+1)\theta}),
    \end{aligned}
\end{equation}
where $A_{\pm,i}$ are invariant under the rotation $\Omega_{\pm}$. See \cite{unp-zamo} for the case of $r=1$. Substituting the expansions into the Q-functions and Y-functions, 
one finds the leading contributions of $\log Q_a$, $\log Y_a$ 
\begin{equation}
\begin{aligned}
     \log Q_{a}\sim&-2(r+1)\sum_{i=1}^{a}\lambda_{i}\theta,\\
     \log Y_{a}\sim&-2(r+1)(\lambda_{a}-\lambda_{a+1})\theta, \quad \theta\to-\infty.
\end{aligned}
\end{equation}
We thus find the coefficients $h_a$ in the boundary conditions \eqref{eq:hk-eps-bdy} are given by 
\begin{equation}
    h_{a}=2(r+1)(\lambda_{a}-\lambda_{a+1}).
\end{equation}
Substituting $h_a$ and $\lambda_i$ into the effective central charge \eqref{eq:ceff}, we find
\begin{equation}\label{eq:ceff from boundary conditions}
        c_{\mathrm{eff}}=r+12\sum_{i=1}^{r+1}\lambda_{i}^{2}=\begin{cases}
1+24P^{2} & r=1\\
r+24p_{2} & \text{even }r\\
r-24p_{2} & \text{odd }r
\end{cases}.
\end{equation}

\section{Large-$\theta$ expansion from WKB side}\label{sec:large theta from WKB}

In this section, we compute the asymptotic behaviors of the Q-functions using the WKB expansion at $\theta\to\infty$, which will be compared with the large-$\theta$ expansion obtained from the TBA side.

We compute the asymptotic behavior of the $\log Q_a$ by using the WKB method. 
The WKB expansions at $\theta\to \infty$ start with the 
ansatz $\psi(y,\theta)=\exp\left(\int_{-\infty}^{y}S(x,\theta)dx\right)$, where $S(y,\theta)=\sum_{n=-1}^{\infty}e^{-n\theta}S_{n}(y)$. Substituting this ansatz into \eqref{ODE r+1 theta P}, we obtain the Ricatti-type equation, 
\be
(-1)^{r+2}[\partial_{y}+S(y,\theta)]^{r}S(y,\theta)+e^{(r+1)\theta}\cosh y+P^{r+1}=0.
\ee
The first two orders are given by 
\begin{equation}
    \begin{aligned}
         S_{-1}^{r+1}=(-1)^{r+1}\cosh y,\quad S_{0}(y)=-\frac{r}{2}\partial_y\log{S_{-1}(y)}.
    \end{aligned}
\end{equation}
Higher-order expansion will depend on the value of $r$. In particular, the $P$ dependence will appear at $S_{r}$. 
At the leading order of large-$\theta$, the WKB solution of \eqref{ODE r+1 theta P} reads 
\begin{equation}
\psi\left(y,\theta\right)\sim N\exp\left(e^{\theta}\int_{-\infty}^{y}(-1)^{\frac{1}{r+1}}(\cosh x)^{\frac{1}{r+1}}dx\right), 
\end{equation}
where the branch should be chosen to match the asymptotic behaviors of $\psi_{\pm}$ given in \eqref{eq:psi-pos} and \eqref{eq:psi-neg}. 
The Q-functions connects $\psi_{-}$ with $\psi_{+}$. Evaluating this connection at certain value of $y$, the asymptotic behavior of $\log Q_1$ can be computed. At leading order, we find $\log Q_1\sim e^{\theta}I$, where the function $I$ is given by 
\begin{equation}\label{eq:I general r}
    I:=\int_{-\infty}^{\infty}(\cosh x)^{\frac{1}{r+1}}-2^{-\frac{1}{r+1}}(e^{-\frac{x}{r+1}}+e^{\frac{x}{r+1}})dx=\frac{\Gamma\left(\frac{1}{2}\right)\Gamma\left(-\frac{1}{2(r+1)}\right)}{\Gamma\left(\frac{r}{2(r+1)}\right)}. 
\end{equation}
 See \cite{Fioravanti:2019vxi} for the case of $r=1$. The function $I$ is nothing but $-q_1$. Other $q_a$ and $m_a$ can be computed by using the constraints \eqref{eq:qk constarins}, \eqref{eq:mk qk constraints}. 
 Higher-order corrections of Q-function can be computed by substituting higher-order terms in the WKB ansatz. Since the detailed form of the higher-order corrections of the WKB ansatz depends on $r$, we will illustrate this in the cases of $r=2,3$. 
 The computation for other values of $r$ can be performed in a similar way.

\subsection{$r=2$}
For later convenience, we shift $\theta$ by $i\pi/3$, which leads to a minus sign before the $e^{3\theta}$ term in the ODE, and thus avoids the $(-1)^{1/3}$ in the solution $\psi_{-}$ \eqref{eq:psi-neg}. We denote the WKB ansatz after the shift by 
$\psi\left(y,\theta+\frac{\pi i}{3}\right)=\exp\left(\int_{-\infty}^{y}\tilde{S}(x,\theta)dx\right)$. Here $\tilde{S}(x,\theta)=\sum_{n=-1}^{\infty}e^{-n\theta}\tilde{S}_{n}(x)$, and $\tilde{S}(y,\theta)=S\left(y,\theta+\frac{\pi i}{r+1}\right)=S(y+i\pi,\theta)$. 
The shifted solution  $\psi_{-,0}\left(\theta+\frac{\pi i}{3}\right)$ can be expanded by 
\begin{equation}\label{eq:psi-exp-r2}
    \begin{aligned}
        \psi_{-,0}\left(\theta+\frac{\pi i}{3}\right)=&W_{(0),0,1}\left(\theta+\frac{\pi i}{3}\right)\psi_{+,-1}\left(\theta+\frac{\pi i}{3}\right)+W_{-1,(0),1}\left(\theta+\frac{\pi i}{3}\right)\psi_{+,0}\left(\theta+\frac{\pi i}{3}\right)\\
&+W_{-1,0,(0)}\left(\theta+\frac{\pi i}{3}\right)\psi_{+,1}\left(\theta+\frac{\pi i}{3}\right).
    \end{aligned}
\end{equation}
Since the Wronskian is independent of  $y$, we can evaluate it at large $y$. At $y\to\infty$, only the third term 
can not be ignored, which behaves as $\psi_{+,1}\left(\theta+\frac{\pi i}{3}\right)\sim-N_{+}e^{-\theta-\frac{y}{3}} e^{2^{-\frac{1}{3}}3e^{\theta+\frac{y}{3}}}$. 
The coefficient of $\psi_{+,1}\left(\theta+\frac{\pi i}{3}\right)$ is nothing but $Q_1 (\theta)$ 
by noting $W_{-1,0,(0)}\left(\theta+\frac{\pi i}{3}\right)=W_{0,1,(0)}(\theta)=-Q_1(\theta)$. 
We thus can compute the Q-function by 
\begin{equation}\label{r2 Q1 psi/psi}
    Q_{1}(\theta)=-\lim_{y\to\infty}\frac{\psi_{-,0}\left(\theta+\frac{\pi i}{3}\right)}{\psi_{+,1}\left(\theta+\frac{\pi i}{3}\right)}.
\end{equation}
The numerator $\psi_{-,0}$ can be extended to $y\to\infty$ as 
\begin{equation}\label{eq:psin0 r2 y to inf}
    \psi_{-,0}\left(\theta+\frac{\pi i}{3}\right)=\tilde{N}_{-}\exp\left[-\frac{3}{\sqrt[3]{2}}e^{\theta-\frac{y}{3}}+\frac{3}{\sqrt[3]{2}}e^{\theta+\frac{y}{3}}+\left(\int_{-\infty}^{y}\tilde{S}(x,\theta)-e^{\theta}\frac{1}{\sqrt[3]{2}}(e^{-\frac{x}{3}}+e^{\frac{x}{3}})dx\right)\right],
\end{equation}
where the branch of $\tilde{S}_{-1}(x)$ in $\tilde{S}(x)$ should be chosen to match the asymptotic behavior of $\psi_{-,0}$ at infinity, see \eqref{eq:psi-neg}. 
The term $\left(e^{\theta+\frac{x}{3}}+e^{\theta-\frac{x}{3}}\right)/\sqrt[3]{2}$ in the integrand is introduced to cancel the divergence of the integral. Substituting the first two orders of $\tilde{S}$, we obtain 
\begin{equation}\label{eq:psin0 r2}
    \begin{aligned}
        \lim_{y\to\infty}\psi_{-,0}\left(\theta+\frac{\pi i}{3}\right)=N_{-}\frac{1}{e^{\theta+\frac{\pi i}{3}+\frac{y}{3}}}\exp\left(\frac{3}{\sqrt[3]{2}}e^{\theta+\frac{y}{3}}+e^{\theta}I+\mathcal{O}\left(e^{-\theta}\right)\right),
    \end{aligned}
\end{equation}
where we have rescaled the normalization factor to $N_-$. 
Substituting \eqref{eq:psin0 r2} and $\psi_{+,1}$ into $Q_1$ \eqref{r2 Q1 psi/psi}, the leading orders of $\log Y_1$ and $\log Q_1$ are found to be
\begin{equation}
    m_1=q_1=-I
    =-\frac{\Gamma(\frac{1}{2})\Gamma(-\frac{1}{6})}{\Gamma(\frac{1}{3})}.
\end{equation}
The higher-order corrections of the Q-function can be computed by using the higher-order terms of the WKB ansatz: 
\begin{equation}
    \begin{aligned}
        \tilde{S}_{1}=&\frac{p_{2}}{3\tilde{S}_{-1}}-\frac{(\partial_{y}\tilde{S}_{-1})^{2}}{\tilde{S}_{-1}^{3}}+\frac{2\partial_{y}^{2}\tilde{S}_{-1}}{3\tilde{S}_{-1}^{2}},\\
        \tilde{S}_{2}=&-\frac{P^{3}}{3\tilde{S}_{-1}^{2}}+\frac{p_{2}}{3}\frac{\partial_{y}\tilde{S}_{-1}}{\tilde{S}_{-1}^{3}}-\frac{2(\partial_{y}\tilde{S}_{-1})^{3}}{\tilde{S}_{-1}^{5}}+\frac{2\partial_{y}\tilde{S}_{-1}\partial_{y}^{2}\tilde{S}_{-1}}{\tilde{S}_{-1}^{4}}-\frac{\partial_{y}^{3}\tilde{S}_{-1}}{3\tilde{S}_{-1}^{3}},\\
        \cdots&,
    \end{aligned}
\end{equation}
where the $P$ dependence first appears at $\tilde{S}_2$. 
Substituting these into $\tilde{S}$ of \eqref{eq:psin0 r2 y to inf}, the higher-order corrections of $\log Q_1$ are given by 
\begin{equation}\label{eq:q-exp-r2-wkb}
    \begin{aligned}
        \log Q_{1}(\theta)=&e^{\theta}I+\int_{-\infty}^{\infty}dy\tilde{S}_{0}+e^{-\theta}\int_{-\infty}^{\infty}dy\tilde{S}_{1}+e^{-2\theta}\int_{-\infty}^{\infty}dy\tilde{S}_{2}+\cdots\\
    =&e^{\theta}\frac{\Gamma\left(\frac{1}{2}\right)\Gamma\left(-\frac{1}{6}\right)}{\Gamma\left(\frac{1}{3}\right)}+0+e^{-\theta}\frac{(1+12p_{2})}{36}\frac{\Gamma\left(\frac{1}{2}\right)\Gamma\left(\frac{1}{6}\right)}{\Gamma\left(\frac{2}{3}\right)}-e^{-2\theta}\frac{4\sqrt{3}P^{3}}{9}\frac{\Gamma\left(\frac{1}{2}\right)\Gamma\left(\frac{7}{6}\right)}{\Gamma\left(\frac{5}{3}\right)}+\cdots,
     \end{aligned}
\end{equation}
where the integrals of $\tilde{S}_1$, $\tilde{S}_2$ should produce $-q_1^{(1)}$, $-q_1^{(2)}$, respectively.
Comparing this with the relation \eqref{eq:q1-ceff}, the effective central charge is given by
\begin{equation}
    c_{{\rm eff}}=-\frac{12m_{1}}{\pi\sqrt{3}}q_{1}^{(1)}=2+24p_{2},
\end{equation}
which reproduces \eqref{eq:ceff from boundary conditions} exactly for $r=2$. The $P$ dependence first appears at the order $e^{-2\theta}$ and the $p_2$ dependence first appears at the order $e^{-\theta}$. However, the analytical calculation of $q_1^{(2)}$ \eqref{eq:q12-r2-TBA} from the TBA is not clear so far. On the other hand, we can solve the TBA equations numerically and test the expansion of $Q_1$ function \eqref{eq:Q-exp-r2-TBA} against the WKB expansion \eqref{eq:q-exp-r2-wkb}.
  In table \ref{tab:tba_wkb-r2}, we compare the coefficients in the large-$\theta$ expansions of $Q_1(\theta)$ obtained from the TBA equations and the WKB expansion, which shows a precise coincidence.

\begin{table}[]
    \centering
    \begin{tabular}{|c|c|c|}\hline
    $n$&   $\big(q^{(n)}_{1}\big)_{\rm WKB}$ & $\big(q^{(n)}_{1}\big)_{\rm TBA}$  \\\hline
    $1$ & $-\frac{\pi\sqrt{3}}{12m_1}\times14$ &$-\frac{\pi\sqrt{3}}{12m_1}\times13.999866$\\\hline
    $2$& $0.05193267$ & $0.05193206$ \\
   \hline
    \end{tabular}    \caption{
 Comparison of the large-$\theta$ expansions of $Q_1(\theta)$ based on the TBA equations and the WKB expansion at $p_2=1/2, P=1/3$ and $r=2$. The TBA equations are solved numerically by using the discretized Fourier transformation with $2^{12}$ points and the cutoff  $(-250,250)$.}
    \label{tab:tba_wkb-r2}
\end{table}

\subsection{$r=3$}

$\psi_{-,0}$ can be expanded by
\begin{equation}
    \begin{aligned}
        \psi_{-,0}=&W_{(0),0,1,2}\psi_{+,-1}+W_{-1,(0),1,2}\psi_{+,0}+W_{-1,0,(0),2}\psi_{+,1}+W_{-1,0,1,(0)}\psi_{+,2}\\
    \end{aligned}
\end{equation}
where only $\psi_{+,2}$ can not be ignored at $y\to\infty$. In a similar way as the previous example, we can evaluate the Q-function at 
$y\to\infty$ by 
\begin{equation}
    Q_{1}=W_{-1,0,1,(0)}=\lim_{y\to\infty}\frac{\psi_{-,0}}{\psi_{+,2}},
\end{equation}
where $\psi_{+,2}\sim-iN_{+}e^{-\frac{3}{2}\theta-\frac{3}{8}y}\exp\left(2^{-\frac{1}{4}}4e^{\theta+\frac{y}{4}}\right)$. The numerator $\psi_{-,0}$ can be evaluated at $y\to\infty$ by
\begin{equation}\label{eq:psin0 r3}
    \lim_{y\to\infty}\psi_{-,0}=N_{-}\frac{1}{\sqrt{e^{3\theta+\frac{3}{4}y}}}\exp\left(2^{-\frac{1}{4}}4e^{\theta+\frac{y}{4}}+e^{\theta}I+\mathcal{O}\left(e^{-\theta}\right)\right),
\end{equation}
Substituting \eqref{eq:psin0 r3} and $\psi_{+,1}$ into $Q_1$, the leading contribution of $\log Q_1$ is given by 
\begin{equation}
    q_1=-I
    =-\frac{\Gamma\left(\frac{1}{2}\right)\Gamma\left(-\frac{1}{8}\right)}{\Gamma\left(\frac{3}{8}\right)}.
\end{equation}
By substituting the higher-order correction of the WKB ansatz
\begin{equation}
\begin{aligned}
    S_{1}=&-\frac{p_{2}}{4S_{-1}}+\frac{5\partial_{y}^{2}S_{-1}}{4S_{-1}^{2}}-\frac{15(\partial_{y}S_{-1})^{2}}{8S_{-1}^{3}},\\S_{2}=&-\frac{p_{3}}{4S_{-1}^{2}}+\frac{-p_{2}\partial_{y}S_{-1}-\frac{5}{2}\partial_{y}^{3}S_{-1}}{4S_{-1}^{3}}+\frac{15\partial_{y}S_{-1}\partial_{y}^{2}S_{-1}}{4S_{-1}^{4}}-\frac{15(\partial_{y}S_{-1})^{3}}{4S_{-1}^{5}},\\S_{3}=&\frac{p_{2}^{2}+8P^{4}}{32S_{-1}^{3}}+\frac{-12p_{3}\partial_{y}S_{-1}-p_{2}\partial_{y}^{2}S_{-1}+\partial_{y}^{4}S_{-1}}{16S_{-1}^{4}}+\frac{75(\partial_{y}S_{-1})^{2}\partial_{y}^{2}S_{-1}}{32S_{-1}^{6}}\\&+\frac{p_{2}(\partial_{y}S_{-1})^{2}-20\partial_{y}^{3}S_{-1}\partial_{y}S_{-1}-5(\partial_{y}^{2}S_{-1})^{2}}{32S_{-1}^{5}}-\frac{225(\partial_{y}S_{-1})^{4}}{128S_{-1}^{7}},\\
    \cdots&,
\end{aligned}
\end{equation}
into $\psi_{-,0}$ \eqref{eq:psin0 r3}, we can compute the higher-order correction of $\log Q_1$, 
\begin{equation}
    \begin{aligned}
        \log Q_{1}=&e^{\theta}I+\int_{-\infty}^{\infty}dy\ S_{0}+e^{-\theta}\int_{-\infty}^{\infty}dy\ S_{1}+e^{-2\theta}\int_{-\infty}^{\infty}dy\ S_{2}+e^{-3\theta}\int_{-\infty}^{\infty}dy\ S_{3}+\cdots\\
    =&e^{\theta}\frac{\Gamma\left(\frac{1}{2}\right)\Gamma\left(-\frac{1}{8}\right)}{\Gamma\left(\frac{3}{8}\right)}+0+e^{-\theta}\frac{\left(1-8p_{2}\right)}{32}\frac{\Gamma\left(\frac{1}{2}\right)\Gamma\left(\frac{1}{8}\right)}{\Gamma\left(\frac{5}{8}\right)}-e^{-2\theta}\frac{p_{3}}{4}\frac{\Gamma\left(\frac{1}{2}\right)\Gamma\left(\frac{1}{4}\right)}{\Gamma\left(\frac{3}{4}\right)}\\&+e^{-3\theta}\frac{\left(80p_{2}\left(4p_{2}-1\right)+2560P^{4}+41\right)}{10240}\frac{\Gamma\left(\frac{1}{2}\right)\Gamma\left(\frac{3}{8}\right)}{\Gamma\left(\frac{7}{8}\right)}+\cdots,
    \end{aligned}
\end{equation}
where the $p_2$, $p_3$ and $P$ dependence first appears at order $e^{-\theta}$, $e^{-2\theta}$ and $e^{-3\theta}$, respectively. Comparing this with the relation \eqref{eq:q1-ceff r3}, the effective central charge is given by
\begin{equation}
    c_{{\rm eff}}=-\frac{6\sqrt{2}m_{1}}{\pi}q_{1}^{(1)}=3-24p_{2},
\end{equation}
where we have used $m_{1}=(2-\sqrt{2})q_{1}$. This reproduces \eqref{eq:ceff from boundary conditions} exactly for $r=3$. In table \ref{tab:tba_wkb-r3}, we compare the coefficients $q_1^{(1)}, q_1^{(2)}$ and $q_1^{(3)}$ computed from the TBA approach and the WKB expansions, which also shows a precise match.

\begin{table}[htp]
    \centering
    \begin{tabular}{|c|c|c|c|c|}\hline
    $n$&   $\big(q^{(n)}_{1}\big)_{\rm WKB}$  & $\big(q^{(n)}_{1}\big)_{\rm TBA}$  \\\hline
    $1$ & $-\frac{\pi}{6\sqrt{2}m_1}\times 51$&$-\frac{\pi}{6\sqrt{2}m_1}\times 51.000020$\\\hline
    $2$ & $-1.3110288$ &$-1.3110287$\\\hline
    $3$ &$-0.61790806$ &$-0.61790783$\\\hline
    \end{tabular}    \caption{
 Comparison of the large-$\theta$ expansions of Q-function based on the TBA equations and the WKB expansion at $p_2=-2, p_3=-1$, $P=1/2$ and $r=3$. The TBA equations are solved numerically by using the discretized Fourier transformation with $2^{12}$ points and the cutoff  $(-250,250)$.}
    \label{tab:tba_wkb-r3}
\end{table}

\section{Conclusion and discussion}

In this work, we have developed the ODE/IM correspondence for the higher-order Mathieu equation arising from the quantum SW curve of the pure $SU(r+1)$ SYM \eqref{ODE r+1 theta P}. Starting from the subdominant solutions of the higher-order Mathieu equation, we constructed the Q- and Y-systems and converted them into the TBA equations. The dependence on the moduli parameters ($P$, $p_a$) is encoded in the boundary conditions of the Y-function at $\theta \to -\infty$, which can be imposed into the TBA by using the Zamolodchikov's trick. Based on these boundary data, we have computed the effective central charge analytically, which also controls the subleading term in the large-$\theta$ expansion of the Y-function. We further compared the large-$\theta$ expansion of the $Q$-function derived from the TBA equations with that obtained from the WKB method. The agreement is exact at subleading order, where the WKB side reproduces the effective central charge, and it persists at higher orders, as confirmed by high-precision numerical tests. 

The framework developed here opens a natural route toward more general quantum SW curves and their associated integrable structures. One immediate direction is to extend the present analysis to generalized higher-order Mathieu equations with potentials of the form $\mu_+e^{\theta/b}+\mu_-e^{\theta b}$, which should be relevant for exploring the integrable structures of the corresponding CFTs. In this paper, we have transformed the quantum SW curve into the higher-order Mathieu equation. It would be worthwhile to revisit the problem by keeping the ODE in its original quantum SW curve form. 
In this formulation, the Y-system and TBA equations will be more complicated due to the shifts of the moduli parameters. 
Such TBA equations are expected to encode the resurgent properties of the quantum periods in the strong coupling of the SW theory. See the case of $r=1$ in \cite{Fioravanti:2019vxi}. 
To extend the TBA into the whole moduli space, one should also consider the wall crossing of the corresponding TBA system. 

It would also be interesting to extend our construction to $SU(r+1)$ quantum SW curve with matter, where regular singularities will be incorporated. Another promising direction is to consider an alternative quantization of the SW curve \eqref{eq:SW curve} by taking $y\to \epsilon \partial_p$. This prescription leads to difference equations rather than ordinary differential equations, and therefore provides another framework in which exact WKB methods and the ODE/IM correspondence may be unified. Related developments can be found in \cite{Grassi:2018bci,DelMonte:2022kxh,Baldino:2023nbc,DelMonte:2024dcr,Gu:2026xgp}.

\subsection*{Acknowledgements}
We would like to thank Davide Fioravanti,  Daniele Gregori, Jie Gu, Katsushi Ito, Yong Li, Marco Rossi, Jingjing Yang, Wenhao Yuan and Hao Zou for useful discussions. H.S. is supported by the National Natural Science Foundation of China (Grant No.12405087), Henan Postdoc Foundation (Grant No.22120055) and the Startup Funding of Zhengzhou University (Grant No.121-35220049, 121-35220581).

\appendix

\section{Normalization}
\label{appendix normalization}
In this section, we will explain why the determinant in $Q_{r+1}$ is equal to $1$. We take the case where $r$ is even as the example; the case where $r$ is odd could be done using the same procedure. Using the expansion \eqref{eq:exp psink even r} of $\psi_{-,a}$, we can write the expansions for $a=0,1,\cdots,r$ as
\begin{equation}
    \left(\begin{array}{c}
\psi_{-,0}\\
\psi_{-,1}\\
\vdots\\
\psi_{-,r}
\end{array}\right)=\left(\begin{array}{cccc}
W_{(0),-\frac{r}{2}+2,\cdots,\frac{r}{2}+1} & W_{-\frac{r}{2}+1,(0),\cdots,\frac{r}{2}+1} & \cdots & W_{-\frac{r}{2}+1,\cdots,\frac{r}{2},(0)}\\
W_{(1),-\frac{r}{2}+2,\cdots,\frac{r}{2}+1} & W_{-\frac{r}{2}+1,(1),\cdots,\frac{r}{2}+1} & \cdots & W_{-\frac{r}{2}+1,\cdots,\frac{r}{2},(1)}\\
\vdots & \vdots & \ddots & \vdots\\
W_{(r),-\frac{r}{2}+2,\cdots,\frac{r}{2}+1} & W_{-\frac{r}{2}+1,(r),\cdots,\frac{r}{2}+1} & \cdots & W_{-\frac{r}{2}+1,\cdots,\frac{r}{2},(r)}
\end{array}\right)\left(\begin{array}{c}
\psi_{+,-\frac{r}{2}+1}\\
\psi_{+,-\frac{r}{2}+2}\\
\vdots\\
\psi_{+,\frac{r}{2}+1}
\end{array}\right).
\end{equation}
Take the Wronskian of the both sides, the matrix equation turns into
\begin{equation}
    W[\psi_{-,0},\psi_{-,1},\cdots,\psi_{-,r}]=\det\left(\cdots\right)W[\psi_{+,-\frac{r}{2}+1},\psi_{+,-\frac{r}{2}+2},\cdots,\psi_{+,\frac{r}{2}+1}],
\end{equation}
where $\left(\cdots\right)$ represents the $(r+1)\times(r+1)$ coefficients matrix and the two Wronskians are $1$ under our normalization. Therefore the determinant of the coefficients matrix $\det\left(\cdots\right)=1$. The equivalence between the determinant of the coefficients matrix and the determinant inside $Q_{r+1}$ can be revealed by expanding the $\psi_{-\frac{r}{2}+1},\psi_{-\frac{r}{2}+2},\cdots$ on the left of $\psi_{-,a}$ in the coefficients using the following expansion,
\begin{equation}
    \psi_{+,\alpha}=\sum_{\sigma=\alpha+1}^{\alpha+r+1}W[\psi_{+,\alpha+1},\cdots,\psi_{+,\sigma-1},\psi_{+,\alpha},\psi_{+,\sigma+1},\cdots,\psi_{+,\alpha+r+1}]\psi_{+,\sigma}.
\end{equation}
We will illustrate this procedure in a simple example, the case of $r=2$.
\subsection{$r=2$}
When $r=2$, using the expansion \eqref{eq:exp psink even r}, $\psi_{-,a}$ can be expanded by
\begin{equation}
    \psi_{-,a}=W_{(a),1,2}\psi_{+,0}+W_{0,(a),2}\psi_{+,1}+W_{0,1,(a)}\psi_{+,2}.
\end{equation}
Substituting this expansion into $W[\psi_{-,0},\psi_{-,1},\psi_{-,2}]$, we have
\begin{equation}\label{eq:det=Q r+1 r=2}
    W[\psi_{-,0},\psi_{-,1},\psi_{-,2}]=\det\left(\begin{array}{ccc}
W_{0,1,(0)} & W_{0,1,(1)} & W_{0,1,(2)}\\
W_{(0),1,2} & W_{(1),1,2} & W_{(2),1,2}\\
W_{0,(0),2} & W_{0,(1),2} & W_{0,(2),2}
\end{array}\right)W[\psi_{+,0},\psi_{+,1},\psi_{+,2}],
\end{equation}
where the elements in the first two rows of the coefficients matrix can already be interpreted as $-Q_1^{[\pm j]}$. For the elements in the 3rd row $W_{0,(a),2}$, we expand the $\psi_{+,0}$ on the left of $\psi_{-,a}$ using the expansion
\begin{equation}
    \psi_{+,0}=W_{0,2,3}\psi_{+,1}+W_{1,0,3}\psi_{+,2}+W_{1,2,0}\psi_{+,3}.
\end{equation}
Substituting the expansion, $W_{0,(a),2}$ turns into
\begin{equation}
    W_{0,(a),2}=W_{0,2,3}W_{1,(a),2}+W_{1,0,3}W_{2,(a),2}+W_{1,2,0}W_{3,(a),2},
\end{equation}
where $W_{1,(a),2}$ coincides with the 2nd row and $W_{2,(a),2}=0$. The only term left is $W_{3,(a),2}=W_{(a),2,3}=-Q_{1}^{[+2-a]}$. Therefore, the determinant in \eqref{eq:det=Q r+1 r=2} is converted into 
\begin{equation}
\det\left(\begin{array}{ccc}
W_{0,1,(0)} & W_{0,1,(1)} & W_{0,1,(2)}\\
W_{(0),1,2} & W_{(1),1,2} & W_{(2),1,2}\\
W_{0,(0),2} & W_{0,(1),2} & W_{0,(2),2}
\end{array}\right)=\det\left(\begin{array}{ccc}
-Q_{1} & -Q_{1}^{[-1]} & -Q_{1}^{[-2]}\\
-Q_{1}^{[+1]} & -Q_{1} & -Q_{1}^{[-1]}\\
-Q_{1}^{[+2]} & -Q_{1}^{[+1]} & -Q_{1}
\end{array}\right), 
\end{equation}
which proves the equivalence between the determinant of the coefficients matrix and the determinant in the $Q_3$.

\bibliography{BIBLIOGRAPHY}
\bibliographystyle{JHEP}

\end{document}